\documentclass[english,aps,pra,amsmath,twocolumn,preprintnumber,superscriptaddress,showpacs,notitlepage]{revtex4-1}
\usepackage[T1]{fontenc}
\usepackage[latin9]{inputenc}
\makeatletter
\usepackage{mathptmx,newtxtext,newtxmath,xspace}
\usepackage{amsbsy,bm,bbold}
\usepackage{graphicx,color,xcolor,epsfig,rotate}
\usepackage{fancyhdr}
\usepackage{gensymb}
\usepackage[colorlinks=true, 
            linkcolor=blue, 
            urlcolor=blue,
            citecolor=blue]{hyperref}
\usepackage{soul}
\allowdisplaybreaks[2]

\newcommand{\beq}{\begin{equation}}   
\newcommand{\eeq}{\end{equation}}
\newcommand{\beqn}{\begin{eqnarray}}   
\newcommand{\eeqn}{\end{eqnarray}}
\newcommand{\pt}{\partial}
\newcommand{\SOthree}{\mbox{SO}(3)}
\newcommand{\SOfour}{\mbox{SO}(4)}
\newcommand{\SU}{\mbox{SU}(2)}

\usepackage[normalem]{ulem}
\newcommand{\stkout}[1]{\ifmmode\text{\sout{\ensuremath{#1}}}\else\sout{#1}\fi}

\definecolor{magenta}{cmyk}{0,1,0,0}


\usepackage{babel}

\begin{document}

\begin{flushright}
FTPI-MINN-18/13, UMN-TH-3724/18
\end{flushright}

\title{Principal Chiral Model in Correlated Electron Systems}


\author{Cristian~D.~Batista}

\affiliation{Department of Physics and Astronomy, The University of Tennessee,
Knoxville, Tennessee 37996, USA}

\affiliation{Quantum Condensed Matter Division and Shull-Wollan Center, Oak Ridge
National Laboratory, Oak Ridge, Tennessee 37831, USA}

\author{Mikhail~Shifman}

\affiliation{William I. Fine Theoretical Physics Institute, University of Minnesota,
Minneapolis, Minnesota 55455, USA}

\author{Zhentao~Wang}
\email{ztwang@utk.edu}
\affiliation{Department of Physics and Astronomy, The University of Tennessee,
Knoxville, Tennessee 37996, USA}

\author{Shang-Shun~Zhang}

\affiliation{Department of Physics and Astronomy, The University of Tennessee,
Knoxville, Tennessee 37996, USA}

\date{\today}
\begin{abstract}
We discuss noncollinear magnetic phenomena whose local order parameter is characterized by more than one spin vector. 
By focusing on the simple cases of 2D triangular and 3D pyrochlore lattices, we demonstrate that their low-energy theories can be described by a one-parametric class of sigma models continuously interpolating between the classical Heisenberg model and the principal chiral model ${\rm Tr} \,\left( \partial_a U \partial_a U^\dagger\right)$ for all $U\in \text{SU}(2)$. The target space can be viewed as a U$(1)$ fibration over the $CP(1)$ space. The 3D version of our model is further generalized to break spatial and spin rotation symmetry $\text{SO}(3) \times \text{SO}(3) \to \text{SO}(3)$.
\end{abstract}

\pacs{~}

\maketitle

The principal chiral model (PCM) is the first term of the model proposed by Skyrme to describe nucleons as topological solitons of the underlying pion field~\cite{Skyrme1961}. The realization of this model in magnetism anticipates the possibility of observing quasispherically symmetric pointlike skyrmions in real materials. These skyrmions are different from the two-dimensional (2D) ``baby'' skyrmion configurations  recently reported in chiral collinear ferromagnets~\cite{Muhlbauer09,Yu10,Nagaosa13}, because noncollinear and collinear magnets have different order parameter manifolds (target spaces): $\SOthree$ and $\text{S}^2$, respectively. The corresponding homotopy groups are $\pi_3[\SOthree]=Z$ and  $\pi_2(\text{S}^2)=Z$, implying that skyrmions of collinear magnets, such as ferromagnets, are pointlike particles in 2D systems (thin films), while skyrmions of noncollinear magnets are pointlike particles in three-dimensional (3D) bulk materials.

Below, we will show that low-energy magnetic phenomena  are described 
by a class of sigma models interpolating between the classical Heisenberg model and the PCM [see \eqref{intr10} with $\beta =0$]. 
The continuous limit of the classical Heisenberg model  is a sigma model with the  $\text{S}^2$ target space, while the PCM 
in the case at hand has the $\text{S}^3$ target space.
The interpolation  corresponds to a continuous deformation of noncollinear magnetic orderings into collinear orderings described by a single vector. We present examples of 2D and 3D lattices, starting from the case of three orthonormal unit vectors and then expanding the theory to a more general situation. 
The low-energy theory  (after integrating out gapped modes) can
be described by virtue of three-component vectors $\bm{e}_\mu$, where the subscript $\mu$ marks distinct vectors ($\mu =1,2,3$) while the superscript $A$ below marks the components in the spin space ($A=1,2,3$). The basis set 
can be chosen as follows~\footnote{Equations (\ref{intr1}) and  (\ref{intr3}) mix the spin index $A$ with the number index $\mu$. Therefore, in what follows, the sums over $A$ and $\mu$ are indistinguishable.}:
\beq
{e}_\mu^A =\delta_\mu^A\,.
\label{intr1}
\eeq
Quantum (gapless) fluctuations rotate the original basis set by the coordinate-dependent matrix 
\beq
{\bm{m}_\mu} (\bm{x})=  R ({\bm{x}})  {\bm e}_\mu, \quad R ({\bm{x}}) \in \SOthree\,.
\label{intr2}
\eeq
If we use the $\SU$ conventions,
\beq
m_\mu^A ({\bm{x}}) =\frac 12 \mbox {Tr}\, \left[U({\bm{x}}) \hat{e}_\mu U^\dagger({\bm{x}}) \sigma^A
\right],  \quad \hat{e}_\mu \equiv {\bm{e}}_\mu \cdot {\bm{\sigma}},
\label{intr3}
\eeq
where $U({\bm{x}})\in \SU$ and ${\bm{\sigma}}$ stands for three Pauli matrices.

The convolution over the spatial derivatives $\pt_a$ ($a=1,2$ for 2D models and 
$a=1,2,3$ for 3D models) reduces to~\footnote{Here and below, if the index $a$ is repeated twice, the sum over $a$ is implied.}
\beq
\pt_a {\bm{m}}_{\mu_0} \cdot \pt_a {\bm{m}}_{\mu_0}
=
4 \left( \sum_{\mu=1,2,3} J_a^{\mu}  J_a^{\mu} \right)-  4J_a^{\mu_0}  J_a^{\mu_0} \,,\\
\eeq
where in the above formula there is no summation over $\mu_0$. 
The current $J_a$ is defined as
\beq
J_a = \left(\partial_a U^\dagger \right)U \equiv \sum_{\mu =1,2,3} i\, J_a^\mu\sigma^\mu ,
\label{pa15}
\eeq
and 
$
J_a^{\mu} = -\frac{i}{2}\, {\rm Tr}\, \left( J_a  \sigma^{\mu}\right).
\label{pa16}
$
Now, if we sum over all three values of $\mu_0=1,2,3$, we arrive at 
\beq
\! \! \!  \! \! \! \sum_{\mu_0=1,2,3} \! \pt_a {\bm{m}}_{\mu_0} \! \cdot \pt_a {\bm{m}}_{\mu_0}
=8  \!\!\sum_{\mu=1,2,3} \! \! J_a^{\mu}  J_a^{\mu} = 4 \,\text{Tr} \,\left( \partial_a U \partial_a U^\dagger\right).
\label{intr7}
\eeq
Up to an overall constant, the last expression is nothing but the PCM, which
has an $\SU\times\SU/Z_2 \cong \SOfour$ chiral symmetry, namely,
$U \to {\cal O}_1 U {\cal O}_2$ with arbitrary ${\cal O}_1, {\cal O}_2 \in \SU$.  The ground state partly breaks the above symmetry. Indeed, if we impose (for definiteness) the boundary condition
$U(\infty)= {\mathbb{1}}$, the ground state becomes $U({\bm x})\equiv {\mathbb{1}}$,  spontaneously breaking the chiral $\SOfour$ symmetry down to a ``diagonal'' $\SOthree$ isospin symmetry corresponding to the requirement   ${\cal O}_2 = {\cal O}_1^\dagger$.

\begin{figure}[t!]
\centering
 \includegraphics[width=0.9\columnwidth]{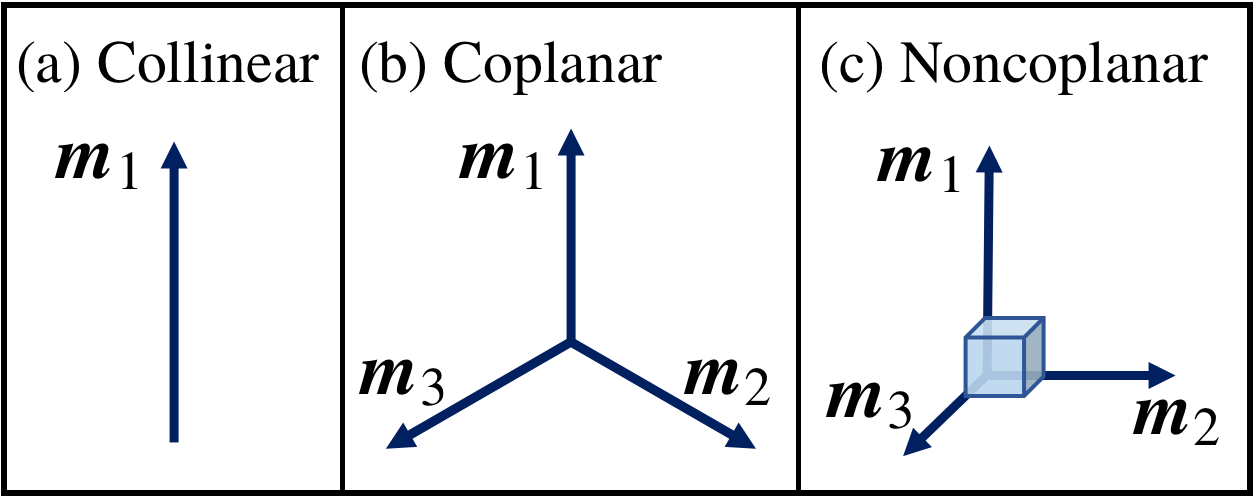}
\caption{Typical cases of  ordered magnets that fall under the general scheme described by Eq.~\eqref{pa24}. 
(a) Collinear ordering ${\bm m}_1={\bm m}_2={\bm m}_3$. (b) Coplanar $120\degree$ ordering of NN Heisenberg antiferromagnet on the TL~\cite{Dombre1989}. (c) Noncoplanar orderings considered in this Letter.
\label{fig:Fig1}}
\end{figure}

The model \eqref{intr7} can be generalized in various directions. For instance, in the left-hand side of \eqref{intr7}
in the sum over $\mu_0$, one can drop one term, say, with $\mu_0=3$.
Then, instead of \eqref{intr7} we will get
\beq
\sum_{\mu_0=1,2} \pt_a {\bm{m}}_{\mu_0} \cdot \pt_a {\bm{m}}_{\mu_0}
=4  \!\!\sum_{\mu=1,2,3} J_a^{\mu}  J_a^{\mu} + 4  J_a^{3}  J_a^{3} \,.
\label{intr8}
\eeq
More generally, one can replace the set \eqref{intr1} by the following set:
\begin{equation}
{\bm{m}}_\mu \cdot {\bm{m}}_\nu = \alpha\,,\quad \mu\neq \nu\,,\quad 0 \leq \alpha \leq  1,\,
\; {\rm and} \;\; \bm{m}_{\nu}^2 =1.
\label{pa24}
\end{equation}
Here $\alpha$ is the cosine of the angle between any pair of the basic vectors. In both cases above and in similar situations,
the Hamiltonian will be generically proportional to 
\beq
{\cal H} \sim \sum_{\mu=1,2,3} J_a^{\mu}  J_a^{\mu} - \beta  J_a^{3}  J_a^{3},
\label{intr10}
\eeq
where $\beta$ is a numerical parameter. For the Heisenberg model  we have $\beta =1$, while  $\beta =0$ for the PCM. 
A similar model was considered in a somewhat different context in the massive $CP(N)$ model for frustrated spin systems~\cite{Azaria1995}.
As we will see below, the latter case is realized by certain noncoplanar magnets, while the intermediate $0 < \beta < 1$ case is realized by the  nearest-neighbor (NN) {\it antiferromagnetic} (AFM) Heisenberg model on the {\it triangular lattice} (TL)~\citep{Dombre1989} (see Fig.~\ref{fig:Fig1}).
There is a whole one-parametric family of models (with $\beta$ neither zero nor one) which interpolates 
between them. The target space of the emerging sigma model is a deformed sphere $\text{S}^3_\beta$ which is topologically equivalent to an $\text{S}^3$, 3D sphere
(nondeformed $\text{S}^3$ corresponds to $\beta = 0$). If $\beta\neq 1$, the target space represents a U(1) fibration over $CP(1)$ space. 
In terms of three angles, the metric of \eqref{intr10} is determined by
\beq
ds^2 \sim  \left\{ \underbrace{d\theta^2 +\cos^2\theta\, d\phi^2}_{ \text{S}^2} +(1-\beta)\underbrace{\left( d\psi + \sin\theta d\phi\right)^2}_{\text{S}^1 \text{ fibration}}
\right\},
\label{intr11}
\eeq
where here and below $\sim$ means proportional; i.e., inessential overall numerical factors are omitted.
It is obvious that for $\beta=1$ we return to $\text{S}^2$, i.e., the Heisenberg model. The scalar curvature corresponding to \eqref{intr11} is constant over the target space and proportional to $\frac 14 (3 +\beta)$. Moreover, $\pi_3 (\text{S}^3_\beta ) = Z$ for all $\beta<1$, implying that skyrmion-type solutions are supported by topology. 
Four-derivative terms are briefly discussed after Eq.~\eqref{eq:27}.

The formulation \eqref{pa24} is physically most transparent in exhibiting the continuous interpolation between the Heisenberg model and PCM in the magnetic phenomena. Indeed, if $\alpha=1$, all three vectors ${\bm{m}}_\mu$ ($\mu=1,2,3$) are collinear and fluctuate as a single vector, while when $\alpha=0$ we return to 
\eqref{intr7} (see Fig.~\ref{fig:Fig1}).

The continuous family of the sigma models presented above (to be revealed below in some magnetic phenomena) plays a special role in mathematical physics. 
First of all, in 2D all these models are integrable and were exactly solved~\cite{Z1,Z2,Z3,Z4,Z5,Z6} by various methods, which makes them rather unique. Second, they possess topologically nontrivial excitations~\footnote{A four-derivative term must be added for IR stabilization in three dimensions.} of the skyrmion type~\cite{Skyrme1961,Z8}. If the 
parameter $\beta$ in \eqref{intr10} is set to zero, the target space is just the round sphere $\text{S}^3$, and the topological excitation is just the standard skyrmion. However,
with increasing $\beta$ (keeping $\beta$ positive), we deform it, and it would be very interesting to trace the evolution of the topological excitation, especially when $\beta$ approaches unity; i.e., the target space becomes close to $\text{S}^2$. 

{\it Two-dimensional model.}---Now we will consider a half-filled Hubbard model~\citep{Gutzwiller1963,Hubbard1963,Kanamori1963,Hubbard1964,Hubbard1964-2,Hubbard1965}
on the TL:
\begin{equation}
\mathcal{H}=-\sum_{\langle ij\rangle}\sum_{\sigma}t_{ij}\left(c_{i\sigma}^{\dagger}c_{j\sigma}+\text{H.c.}\right)+U\sum_{i}n_{i\uparrow}n_{i\downarrow},
\end{equation}
where $t_{ij}$ are the hopping amplitudes up to the third nearest neighbor
$\{t_{1},t_{2},t_{3}\}$ and $U$ is the on-site electron repulsion. Here we will assume that $|t_2|, |t_3| < |t_1|$.
This model can be realized in the adatom system on a semiconductor surface Sn/Si(111)-($\sqrt{3} \times \sqrt{3}$)~\cite{Li13} or in transition metal dichalcogenide moir{\' e} bands~\cite{WuFengcheng2018}.

In the large-$U$ limit, $\mathcal{H}$ can be reduced to an effective $S=1/2$ spin model $\tilde{\mathcal{H}}$ by expanding in the small $t_{\mu}/U$ ratio ($\mu=1,2,3$).
To fourth order in $t_{1}$ and second order in $t_2$ and $t_3$,  
$
\tilde{\mathcal{H}}=\tilde{\mathcal{H}}_{J}+\tilde{\mathcal{H}}_{K}
$  
contains both AFM Heisenberg
and plaquette exchange terms~\citep{Takahashi1977,Roger1983}:
\begin{align}
\tilde{\mathcal{H}}_{J} & = \sum_{\langle ij\rangle}J_{ij}\bm{S}_{i}\cdot\bm{S}_{j},\;\;
\tilde{\mathcal{H}}_{K} = K\sum_{\langle ijkl\rangle}\big[\left(\bm{S}_{i}\cdot\bm{S}_{j}\right)\left(\bm{S}_{k}\cdot\bm{S}_{l}\right) \nonumber \\
&\quad + \left(\bm{S}_{i}\cdot\bm{S}_{k}\right)\left(\bm{S}_{j}\cdot\bm{S}_{l}\right) 
 -\left(\bm{S}_{i}\cdot\bm{S}_{l}\right)\left(\bm{S}_{j}\cdot\bm{S}_{k}\right)\big].
\end{align}
The sum in $\tilde{\mathcal{H}}_{J}$ runs over all the possible bonds $ij$ with $J_{ij}=J_{1}=4t_{1}^{2}/U-28t_{1}^{4}/U^{3}$
for the NN, $J_{ij}=J_{2}=4t_{2}^{2}/U+4t_{1}^{4}/U^{3}$ for the next NN (NNN),
and $J_{ij}=J_{3}=4t_{3}^{2}/U+4t_{1}^{4}/U^{3}$ for the third NN.
The sum in $\tilde{\mathcal{H}}_{K}$ runs over all the smallest parallelograms, with
$\{j,k\}$ connecting the NNN and $K=80t_{1}^{4}/U^{3}$.

Assuming that $U$ is large enough for the Heisenberg term to be dominant, the ordering wave vectors in the classical limit are obtained by minimizing the classical energy of $\tilde{\mathcal{H}}_{J}$,
$
E_{\text{cl}}=\sum_{\eta=1,2,3}\sum_{\bm{\delta}_{\eta}}J_{\eta}S^2\cos\bm{k}\cdot\bm{\delta}_{\eta}
$,
where $\bm{\delta}_{\eta}$ runs over the relative positions of the
$\eta$th NN. 
In this Letter, we will  focus on the parameter space where the ordering wave vectors are the ${\bm M}$ points (half of reciprocal lattice vectors along the [1/2, $\pm \sqrt{3}/2$],  and [0,-1] directions) of the Brillouin
zone of the TL, which will be denoted by $\bm{Q}_{\mu}$ ($\mu=1,2,3$)~\cite{Kubo1997,Momoi1997,Martin2008}.

\begin{figure}[!tbp]
	\centering
	\includegraphics[width=0.95\columnwidth]{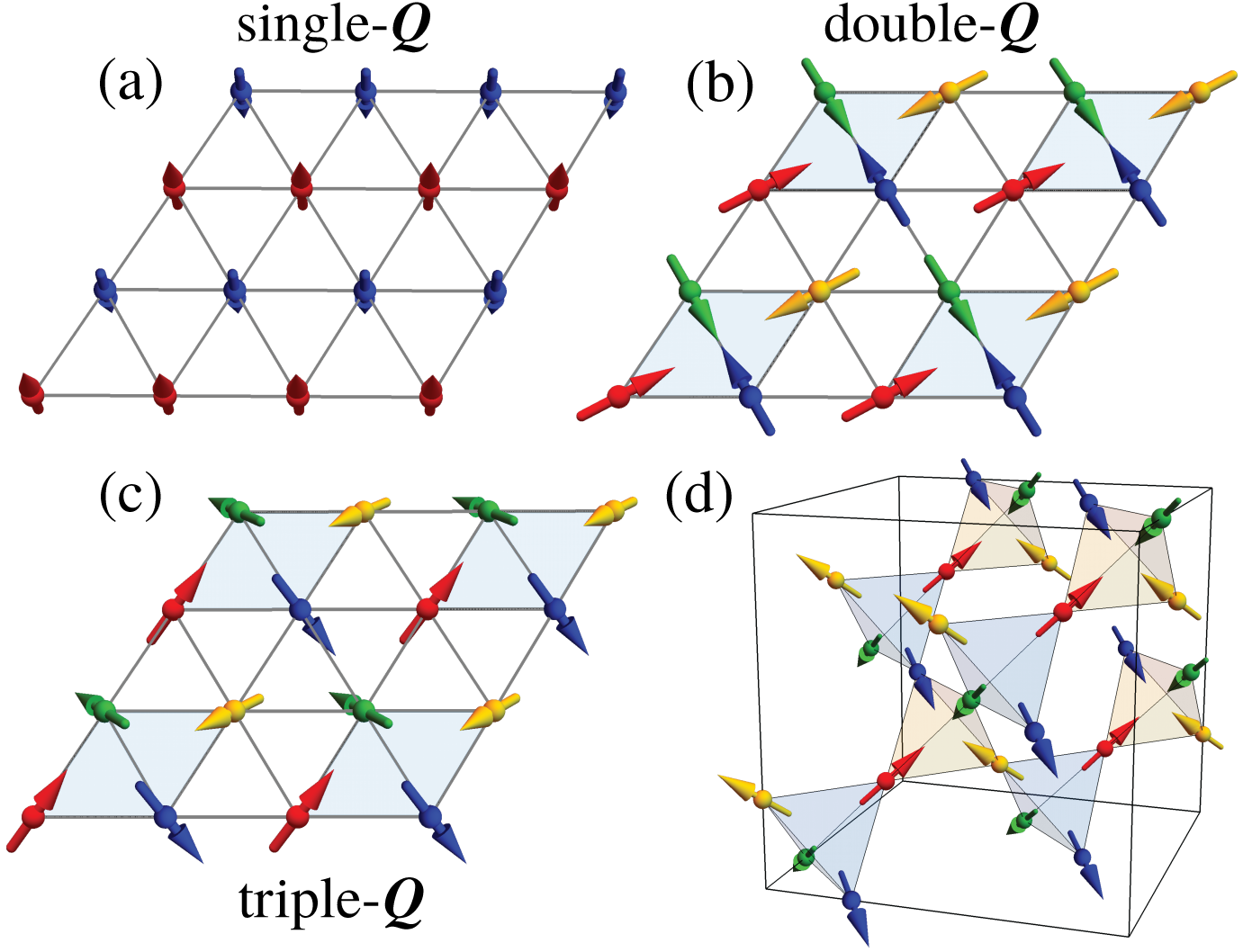}
	\caption{(a) Single-$\bm{Q}$ collinear, (b) double-$\bm{Q}$ coplanar, and (c) triple-$\bm{Q}$ noncoplanar ordering on the TL. (d) All-in--all-out triple-$\bm{Q}$ ordering on the PL.  \label{fig:Fig2}}
\end{figure}

The classical spin configurations that minimize $\tilde{\mathcal{H}}_{J}$ are
\begin{equation}\label{gsc}
\bm{S}_{\bm{r}} = S\sum_{\mu} {\bm s}_{\mu}\cos\left(\bm{Q}_{\mu}\cdot\bm{r}\right)
\end{equation}
with $\bm{s}_{\mu} \cdot \bm{s}_{\nu} = 0$ for  $\mu \neq \nu$ and
$\sum_{\mu}   \bm{s}_{\mu} \cdot \bm{s}_{\mu}=1$. This classical ground state parametrization includes single-, double-, and triple-${\bm Q}$ orderings.
This accidental degeneracy is removed by quantum fluctuations and by the addition of $\tilde{\mathcal{H}}_{K}$.
Quantum fluctuations (order by disorder)~\citep{villain80,Henley89,Chandra1990} favor the collinear single-$\bm{Q}$ ordering, 
\begin{equation}
\bm{S}_{\bm{r}}= S  {\bm u}_{\mu} \cos\left(\bm{Q}_{\mu}\cdot\bm{r}\right),
\label{SQ}
\end{equation}
depicted in Fig.~\ref{fig:Fig2}(a)  [${\bm u}_{\mu}$ is a unit vector with arbitrary direction because of the
SU(2) invariance of ${\cal H}$]. In contrast, $\tilde{\mathcal{H}}_{K}$ favors the noncoplanar triple-$\bm{Q}$  ordering~\citep{Kubo1997,Momoi1997,Martin2008,Kato2010}
\begin{equation}
\bm{S}_{\bm{r}} = \frac{S}{\sqrt{3}} \sum_{\mu} {\bm u}_{\mu} \cos\left(\bm{Q}_{\mu}\cdot\bm{r}\right),
\label{TQ}
\end{equation}
shown in Fig.~\ref{fig:Fig2}(c). Note that ${\bm u}_{\mu} \cdot {\bm u}_{\nu} = \delta_{\mu \nu}$. The triple-$\bm{Q}$ state breaks not only the continuous $\SU$ symmetry
but also the discrete chiral symmetry with a local order parameter,
\begin{equation}
\bm{S}_{j}\cdot\left(\bm{S}_{k}\times\bm{S}_{l}\right) \neq 0,
\label{chiral}
\end{equation}
associated with the noncoplanar nature of this triple-$\bm{Q}$ ordering.
Here $\{j,k,l\}$ denote the three sites in each smallest triangle, and we adopt the
anticlockwise convention for circulation ($j\rightarrow k\rightarrow l$). Indeed, the triple-$\bm{Q}$ state has {\it uniform scalar chirality} (orbital ferromagnetism)~\cite{Bulaevskii08},
whose sign is fixed by the sign of $\bm{u}_{1}\cdot\left(\bm{u}_{2}\times\bm{u}_{3}\right)$ in Eq.~\eqref{TQ}.
The single-$\bm{Q}$ ordering \eqref{SQ} is stabilized by quantum spin fluctuations, which are dominant for large enough $U/|t_{ij}|$; while the triple-$\bm{Q}$ ordering \eqref{TQ}  is stabilized by charge fluctuations below some critical value of $U/|t_{ij}|$. For completeness, Fig.~\ref{fig:Fig2}(b) also shows the coplanar double-${\bm Q}$ ordering that is obtained when only one out of the three ${\bm s}_{\mu}$ vectors is equal to zero in Eq.~\eqref{gsc}.

We next derive an effective field theory for the triple-$\bm{Q}$ ordering.
The local $\SOthree$ order parameter is a rotation matrix instead of the unit vector that describes collinear (single-${\bm Q}$)  ordering.
The relevant homotopy groups are~\citep{Dombre1989}
\begin{equation}
\pi_{1}\left[\text{SO}(3)\right]=Z_{2},\quad\pi_{2}\left[\text{SO}(3)\right]=0,\quad\pi_{3}\left[\text{SO}(3)\right]=Z.
\end{equation}

The normalized spin orientation on the four sublattices can be parametrized as
\begin{subequations}\label{op1}
\begin{align}
\hat{\Omega}_{j} & =\frac{R\left[\bm{n}_{j}^{0}+\bm{n}_{j}^{b}+\bm{L}\right]}{\sqrt{\left|\bm{n}_{j}^{0}+\bm{n}_{j}^{b}+\bm{L}\right|^{2}}},\quad j=\{1,2,3,4\},\\
\bm{n}_{j}^{0} & \equiv\sum_{\mu}\bm{e}_{\mu}\chi_{\mu j},\quad\bm{n}_{j}^{b}\equiv\sum_{\mu}\bm{e}_{\mu}b_{\mu}\chi_{\mu j},
\label{op2}
\end{align}
\end{subequations}
where $\bm{L}$ determines the net magnetization of each unit cell,
$R$ is an $\text{SO}(3)$ matrix describing the global spin rotations of the triad of ${\bm s}_{\mu}$ vectors,
and $b_{\mu}$ represents magnitude fluctuations of the ${\bm s}_{\mu}$ fields. The mutually orthogonal unit vectors $\bm{e}_{\mu}$ are chosen to point along the Cartesian axes  ($\bm{e}_{1}=\hat{x}$, $\bm{e}_{2}=\hat{y}$,
$\bm{e}_{3}=\hat{z}$), and the matrix $\chi$ is
\begin{equation}
\chi=\frac{1}{\sqrt{3}}\begin{pmatrix}1 & 1 & -1 & -1\\
1 & -1 & 1 & -1\\
1 & -1 & -1 & 1
\end{pmatrix}.
\end{equation}
The vector ${\bm b}$ belongs to 
$1/4$ of the sphere S$^2$,
of radius $\sqrt{3}$  centered at ${\bm c}= -(1,1,1)$: $({\bm b}-{\bm c})^2=3$.
The ground space manifold of the classical limit of $\tilde{\mathcal{H}}_{J}$, defined by Eq.~\eqref{gsc}, is obtained by setting ${\bm L}=0$ in Eq.~\eqref{op1} and forcing the fields $R$ and ${\bm b}$ to be uniform.
Note that the ${\bm b}$ fields continuously connect  the triple-${\bm Q}$ ordering of \eqref{TQ} with double- and single-${\bm Q}$ ordering \eqref{SQ}. These fluctuations are massless for the pure classical $\tilde{\mathcal{H}}_{J}$ model, whose  order parameter manifold becomes $\SOthree\times$S$^2/Z_4$. As we show below,  $\tilde{\mathcal{H}}_{K}$ makes the ${\bm b}$ field massive, reducing the order parameter manifold to $\SOthree \times Z_2$ [$Z_2$ associated with the chiral symmetry in \eqref{chiral}].


To first order in $\bm{L}$ and $\bm{b}$, we have
\begin{equation}
\hat{\Omega}_{j}\approx R\left\{ \bm{n}_{j}^{0}+\bm{n}_{j}^{b}+\left[\bm{L}-\left(\bm{n}_{j}^{0}\cdot\bm{L}\right)\bm{n}_{j}^{0}\right]\right\},\label{eq:Omega_1st_order}
\end{equation}
and the local magnetization is $\frac{2}{3}R\bm{L}$.
The continuum limit is obtained by replacing $\bm{S}_{i}$ with $S\hat{\Omega}_{i}$
in $\tilde{\mathcal{H}}$ and implementing the gradient expansion 
\begin{equation}
\hat{\Omega}_{j}(\bm{r}+\bm{\delta})\approx\hat{\Omega}_{j}(\bm{r})+\left(\bm{\delta}\cdot\nabla\right)\hat{\Omega}_{j}(\bm{r})+\frac{1}{2}\left(\bm{\delta}\cdot\nabla\right)^{2}\hat{\Omega}_{j}(\bm{r}).\label{eq:Taylor}
\end{equation}
Here $1\leq j\leq4$ is a sublattice index, and $\bm{r}$ is the real space coordinate. Because  the triple-$\bm{Q}$ ordering conserves parity, the Hamiltonian contains only even powers of spatial derivatives.



Replacing Eq.~\eqref{eq:Taylor} into $\tilde{\cal H}_{J}[\hat{\Omega}]$
and using Eq.~\eqref{eq:Omega_1st_order}, we  obtain 
$\tilde{\cal H}_{J}[\hat{\Omega}]$ in terms of the $R$, $\bm{b}$,
and $\bm{L}$ fields. Our key observation 
is that the continuum limit of the classical Hamiltonian is 
\begin{align}
\tilde{\cal H}_{J}[\hat{\Omega}] & \approx\frac{S^{2}\left(J_{2}-2J_{3}\right)}{\sqrt{3}}
\sum_{\mu_0=1,2,3}\int d^{2}r\,\left[\left(\partial_{a}R\bm{e}_{\mu_0}\right) \cdot 
\left(\partial_{a}R\bm{e}_{\mu_0}\right)\right]\nonumber \\
 & \quad+\frac{128S^{2}J_{2}}{9\sqrt{3}a_0^{2}}\int d^{2}r\,\bm{L}^{2},
 \label{hconttr}
\end{align}
along the surface $J_{1}=3J_{2}$ ($a_0$ is the lattice constant). The $\bm{L}$ mode is 
massive at this level, while the $\bm{b}$ mode becomes massive upon including the four-spin  interactions:
\begin{equation}
\frac{4S^{4}K}{3\sqrt{3} a_0^2}\int d^{2}r\,\left(\bm{b}^{2}+\frac{2}{3}\sum_{\mu<\nu}\left(b_{\mu}-b_{\nu}\right)^{2}\right).
\end{equation}

Based on our previous discussion, the low-energy sector of $\tilde{\cal H}_{J}+\tilde{\cal H}_{K}$ provides a 2D realization of 
the PCM (i.e., it has an emergent $[\SU \times \SU ]/Z_2 \simeq \SOfour$ chiral symmetry) on the fine-tuned parameter surface $J_{1}=3J_{2}$. In terms of the original lattice model, this emergent symmetry corresponds to independent 
invariance under global spin rotations (present in the microscopic model) and orbital transformations connecting different sublattices, which are not present in the microscopic model.

Our next step is to show that fine-tuning can be avoided in magnets with cubic symmetry at the expense of adding ``compasslike'' terms to the PCM, which couple the spatial and spin variables but preserve the overall rotational symmetry in the long wavelength limit (cubic anisotropy appears only to fourth order in the momentum of the excitations).



{\it Three-dimensional case.}---Now we move on to magnetic orderings on the pyrochlore lattice (PL)~\cite{Chern11}.
We will again consider a spin Hamiltonian $\mathcal{H}_{P}=\mathcal{H}_{H} + \mathcal{H}_{B}$ that is the sum of an AFM Heisenberg model 
with NN, NNN, and third-NN exchange constants $\{J_1,J_2,J_3\}$ and a biquadratic interaction between NN:
\begin{equation}
\mathcal{H}_{H} =  \sum_{\langle i j\rangle} J_{ij} \bm{S}_i \cdot \bm{S}_j,
\;\;\;
\mathcal{H}_{B} =  K \sum_{\langle i j\rangle}   (\bm{S}_i \cdot \bm{S}_j)^2.
\end{equation}
Like in the previous case, the classical ground state spin configurations of $\mathcal{H}_{H}$ are given by Eq.~\eqref{gsc}. 
The difference is that the three ordering wave vectors are now 3D reciprocal lattice vectors  $\bm{Q}_{\mu} = \sqrt{2} \pi \bm{e}_{\mu}$
(our unit of length is the distance between NN sites). In particular, the triple-${\bm Q}$ magnetic ordering now corresponds to
the ``all-in--all-out'' configuration depicted in Fig.~\ref{fig:Fig2}(d). Once again, this configuration can be continuously connected with the coplanar double-${\bm Q}$ and
collinear single-${\bm Q}$ by changing the relative magnitudes of the vectors ${\bm s}_{\mu}$ in Eq.~\eqref{gsc} (rotations of the $\bm{b}$ vector).
Similarly to the previous case, the role of the biquadratic interaction $\mathcal{H}_{B}$ is to select the triple-$\bm{Q}$ ordering, i.e., make the $\bm{b}$ fluctuations massive.

Given that the structure of the local order parameter is identical to the 2D case, we can follow a similar procedure to  derive an effective field theory around the triple-${\bm Q}$ all-in--all-out ordered state. In this case, we get
\begin{align}
     {\mathcal H}_{H}[\hat{\Omega}] & \approx\sum_{\mu\nu ab}\int
     d^{3}r\,\bigg[A\delta_{ab}\delta_{\mu\nu}+B_{1}\delta_{\mu \nu}\delta_{a \mu}\delta_{b\nu}\nonumber \\
      & +B_{2}\left(1-\delta_{\mu \nu}\right)\left(\delta_{a\mu}\delta_{b \nu}+\delta_{b\mu}\delta_{a\nu}\right)\bigg]\partial_{a}R\bm{e}_{\mu}\cdot\partial_{b}R\bm{e}_{\nu} \nonumber \\
      & \quad+\frac{16 S^{2} (J_1+2J_2)}{9 \sqrt{2}a_0^{3}}\int d^{3}r\,\bm{L}^{2},  \label{cb26}
      \end{align}
with $\mu \nu a b=\{ 1,2,3 \}$,
$A=\frac{\sqrt{2}S^{2}(J_2-J_3)}{3a_0}$,
$B_{1}=\frac{S^{2}\left(J_{1}-6J_{2}\right)}{6\sqrt{2}a_0}$, and
$B_{2}=\frac{S^{2}\left(J_{1}-2J_{2}-4J_3\right)}{12\sqrt{2}a_0}$. 
Similar to the 2D case, the $\bm{b}$ mode becomes massive upon the inclusion of biquadratic interactions~\footnote{In the spin coherent state path integral formalism, the classical energy of the biquadratic spin interactions will be further renormalized by replacing $S^4\to S^2(S-\tfrac{1}{2})^2$.}:
\begin{equation}
\frac{\sqrt{2}K}{3a_0^3} S^2 \left( S-\frac{1}{2}\right)^2 \int d^{3}r\,\left[\bm{b}^{2}+\frac{2}{3}\sum_{\mu<\nu}\left(b_{\mu}-b_{\nu}\right)^{2}\right].
\end{equation}

Proceeding in the same way as above, we get the zero-mode Hamiltonian 
\begin{equation}
\!\!\!\!\!\!\!
{\mathcal H}_{H} \approx   f_0 \! \sum_{\mu}  J^\mu_a J^\mu_a  \, + f_1 \! \sum_{a}\, J^a_a  J^a_a
+ f_2
\sum_{a \mu } (J^a_{\mu} J^{\mu}_a + J^a_{a} J^{\mu}_{\mu}),
\label{eq:27}
\end{equation}
with $f_0=B_1+2(A+B_2)$, $f_1=2B_2 -B_1$, and $f_2=-2B_2$.
%
Were the coefficients $B_{1,2}$ to vanish, nothing would change in the above theoretical framework,
except the number of spatial coordinates must be 3. The terms proportional to $B_{1,2}$ mix  the coordinate and spin indices, reducing the overall coordinate-isospin symmetry from spherical to cubic~\footnote{The model is invariant under a simultaneous cubic group transformation in coordinate and isospin space.}. Nevertheless,  given that cubic anisotropy appears only to fourth order in the momentum of the quasiparticles, the low-energy excitation spectrum  of Eq.~\eqref{cb26} remains
identical to the one of the PCM (a degenerate triplet of Goldstone modes with renormalized velocity), implying that the inclusion of the $B_1$ and $B_2$ still preserved the  overall rotational symmetry in the long wavelength limit.

A natural consequence of our derivation is that stable skyrmion configurations can be generically induced in  noncollinear magnets. The stability condition
of the zero-angular-momentum mode
is granted by the four-derivative terms in the gradient expansion (not included in our derivation), which are always present in real magnets and play the role of the Lorentz invariant quartic term in Skyrme's model~\footnote{A discussion of other quartic terms in the context of QCD can be found, for example, in ref.~\cite{Zahed1986}.}. 
This can be verified by splitting the energy into the contributions coming from the two- and four-derivative terms, $E=E_2+E_4$, and applying a scale transformation: ${\bm x} \to \lambda {\bm x}$~\cite{Manton_book}. The energy then becomes $e(\lambda)=E_2/\lambda + \lambda E_4$,
implying that there is an optimal skyrmion size (the one satisfying $E_2=E_4$) and that this size is of the order of $\sqrt{g_4/g_2}$,
where $g_2$ ($g_4$) is the effective coupling constant in front of the two- (four-) derivative term. It is then clear that $g_2$  (stiffness of the nonlinear sigma model) must become much smaller than $g_4$ for the skyrmion radius to be much bigger than the lattice parameter $\sqrt{g_4/g_2} \gg a_0$. This condition can be achieved in the proximity of a Lifshitz point (commensurate to incommensurate transition).  This point can be reached by increasing $J_3$ in the two models considered in this Letter [e.g., for the triangular lattice case, the Lifshitz point is at $J_3=J_2/2$ according to Eq.~\eqref{hconttr}].

In summary, noncollinear magnets can produce locally stable 3D skyrmions under quite general conditions. Moreover, like in the case of the 2D baby skyrmions~\cite{Bogdanov89,Leonov15,Hayami16,batista16,Ozawa17,Lin18}, 3D skyrmion crystals are also expected to occur beyond the Lifshitz  transition. These 3D skyrmion crystals are the magnetic counterpart of the neutron crystal solutions found by Klebanov in the context of neutron stars~\cite{Klebanov85}.

M. S. is grateful to Andrei Losev and Paul Wiegmann for useful discussions. The work of M. S.  is supported in part by Department of Energy Grant No. DE-SC0011842. 
C. D. B., Z. W., and S.-S. Z. are supported by funding from the Lincoln Chair of Excellence in Physics.

\bibliographystyle{apsrev4-1}
\bibliography{ref}

\end{document}